\documentclass{aastex631}

\usepackage{multirow}
\usepackage{newtxtext,newtxmath}
\usepackage{soul}

\usepackage[colorinlistoftodos]{todonotes}
\usepackage{color, colortbl, xcolor}
\definecolor{Red}{rgb}{0.9,0,0}
\definecolor{Blue}{rgb}{0,0,0.9}
\definecolor{Green}{rgb}{0,0.5,0}
\definecolor{Black}{rgb}{0,0,0}

\begin{document}

\title{Size and Shape of Jupiter Trojan (2207) Antenor from Stellar Occultations}

\shorttitle{Stellar occultation by Jupiter Trojan (2207) Antenor}
\shortauthors{Ferreira, F. S., Camargo, J. I. B., Morgado, B. E. et al.}

\author[0000-0002-5879-2173]{F. S. Ferreira}
\affiliation{Observatório Nacional/MCTI, R. General José Cristino 77, CEP 20921-400 Rio de Janeiro - RJ, Brazil}
\affiliation{Laboratório Interinstitucional de e-Astronomia - LIneA - and INCT do e-Universo. Av. Pastor Martin Luther King Jr, 126 Del Castilho, Nova América Offices, Torre 3000/sala 817 CEP: 20765-000, Brazil}
\correspondingauthor{F. S. Ferreira}
\email{felipheferreira@on.br}

\author[0000-0002-1642-4065]{J. I. B. Camargo}
\affiliation{Observatório Nacional/MCTI, R. General José Cristino 77, CEP 20921-400 Rio de Janeiro - RJ, Brazil}
\affiliation{Laboratório Interinstitucional de e-Astronomia - LIneA - and INCT do e-Universo. Av. Pastor Martin Luther King Jr, 126 Del Castilho, Nova América Offices, Torre 3000/sala 817 CEP: 20765-000, Brazil}

\author[0000-0003-0088-1808]{B. E. Morgado}
\affiliation{Universidade Federal do Rio de Janeiro - Observatório do Valongo, Ladeira Pedro Antônio 43, CEP 20.080-090 Rio de Janeiro - RJ, Brazil}
\affiliation{Laboratório Interinstitucional de e-Astronomia - LIneA - and INCT do e-Universo. Av. Pastor Martin Luther King Jr, 126 Del Castilho, Nova América Offices, Torre 3000/sala 817 CEP: 20765-000, Brazil}


\author[0000-0002-3319-4548]{V. F. Peixoto}
\affiliation{Universidade Federal do Rio de Janeiro - Observatório do Valongo, Ladeira Pedro Antônio 43, CEP 20.080-090 Rio de Janeiro - RJ, Brazil}
\affiliation{Observatório Nacional/MCTI, R. General José Cristino 77, CEP 20921-400 Rio de Janeiro - RJ, Brazil}
\affiliation{Laboratório Interinstitucional de e-Astronomia - LIneA - and INCT do e-Universo. Av. Pastor Martin Luther King Jr, 126 Del Castilho, Nova América Offices, Torre 3000/sala 817 CEP: 20765-000, Brazil}

\author[0000-0002-2193-8204]{J. Desmars}
\affiliation{IMCCE, Observatoire de Paris, PSL Université, Sorbonne Université, Université Lille 1, CNRS UMR 8028, 77 avenue Denfert-Rochereau 75014, Paris, France}
\affiliation{Institut Polytechnique des Sciences Avancées IPSA, 63 boulevard de Brandebourg, 94200 Ivry-sur-Seine, France}

\author[0000-0003-2311-2438]{F. Braga-Ribas}
\affiliation{Federal University of Technology - Paraná (UTFPR / DAFIS), Rua Sete de Setembro, 3165, CEP 80230-901, Curitiba, PR, Brazil}
\affiliation{Observatório Nacional/MCTI, R. General José Cristino 77, CEP 20921-400 Rio de Janeiro - RJ, Brazil}
\affiliation{Laboratório Interinstitucional de e-Astronomia - LIneA - and INCT do e-Universo. Av. Pastor Martin Luther King Jr, 126 Del Castilho, Nova América Offices, Torre 3000/sala 817 CEP: 20765-000, Brazil}

\author[0000-0002-3362-2127]{A. R. Gomes-Júnior}
\affiliation{Federal University of Uberlândia (UFU), Physics Institute, Av. João Naves de Ávila 2121, Uberlândia, MG 38408-100, Brazil}
\affiliation{Laboratório Interinstitucional de e-Astronomia - LIneA - and INCT do e-Universo. Av. Pastor Martin Luther King Jr, 126 Del Castilho, Nova América Offices, Torre 3000/sala 817 CEP: 20765-000, Brazil}

\author[0000-0003-1995-0842]{B. Sicardy}
\affiliation{IMCCE, Observatoire de Paris, PSL Université, Sorbonne Université, Université Lille 1, CNRS UMR 8028, 77 avenue Denfert-Rochereau 75014, Paris, France}

\author[0000-0002-8690-2413]{J. L. Ortiz}
\affiliation{Instituto de Astrofísica de Andalucía (IAA-CSIC), Glorieta de la Astronomía s/n. 18008-Granada, Spain}

\author[0000-0003-1690-5704]{R. Vieira-Martins}
\affiliation{Observatório Nacional/MCTI, R. General José Cristino 77, CEP 20921-400 Rio de Janeiro - RJ, Brazil}
\affiliation{Laboratório Interinstitucional de e-Astronomia - LIneA - and INCT do e-Universo. Av. Pastor Martin Luther King Jr, 126 Del Castilho, Nova América Offices, Torre 3000/sala 817 CEP: 20765-000, Brazil}


\author[0000-0003-1000-8113]{C. L. Pereira}
\affiliation{Observatório Nacional/MCTI, R. General José Cristino 77, CEP 20921-400 Rio de Janeiro - RJ, Brazil}
\affiliation{Laboratório Interinstitucional de e-Astronomia - LIneA - and INCT do e-Universo. Av. Pastor Martin Luther King Jr, 126 Del Castilho, Nova América Offices, Torre 3000/sala 817 CEP: 20765-000, Brazil}

\author[0000-0003-3433-6269]{L. Liberato}
\affiliation{Université Côte d'Azur, Observatoire de la Côte d'Azur, CNRS, Laboratoire Lagrange, Bd de l'Observatoire, CS 34229, 06304, Nice Cedex 4, France}
\affiliation{UNESP - São Paulo State University, Grupo de Dinâmica Orbital e Planetologia, CEP 12516-410, Guaratinguetá, SP, Brazil}


\author[0000-0002-8211-0777]{M. Assafin}
\affiliation{Universidade Federal do Rio de Janeiro - Observatório do Valongo, Ladeira Pedro Antônio 43, CEP 20.080-090 Rio de Janeiro - RJ, Brazil}
\affiliation{Laboratório Interinstitucional de e-Astronomia - LIneA - and INCT do e-Universo. Av. Pastor Martin Luther King Jr, 126 Del Castilho, Nova América Offices, Torre 3000/sala 817 CEP: 20765-000, Brazil}

\author[0000-0001-8641-0796]{Y. Kilic}
\affiliation{TÜBITAK National Observatory, Akdeniz University Campus, Antalya 07058, Turkey}
\affiliation{Instituto de Astrofísica de Andalucía (IAA-CSIC), Glorieta de la Astronomía s/n. 18008-Granada, Spain}

\author[0000-0003-3452-1114]{R. C. Boufleur}
\affiliation{Observatório Nacional/MCTI, R. General José Cristino 77, CEP 20921-400 Rio de Janeiro - RJ, Brazil}
\affiliation{Laboratório Interinstitucional de e-Astronomia - LIneA - and INCT do e-Universo. Av. Pastor Martin Luther King Jr, 126 Del Castilho, Nova América Offices, Torre 3000/sala 817 CEP: 20765-000, Brazil}

\author[0000-0002-6085-3182]{F. L. Rommel}
\affiliation{Florida Space Institute, UCF, 12354, Research Parkway, Partnership 1 building, Room 211, Orlando, USA}
\affiliation{Federal University of Technology - Paraná (UTFPR / DAFIS), Rua Sete de Setembro, 3165, CEP 80230-901, Curitiba, PR, Brazil}
\affiliation{Laboratório Interinstitucional de e-Astronomia - LIneA - and INCT do e-Universo. Av. Pastor Martin Luther King Jr, 126 Del Castilho, Nova América Offices, Torre 3000/sala 817 CEP: 20765-000, Brazil}

\author[0000-0002-4106-476X]{G. Benedetti-Rossi}
\affiliation{Laboratório Interinstitucional de e-Astronomia - LIneA - and INCT do e-Universo. Av. Pastor Martin Luther King Jr, 126 Del Castilho, Nova América Offices, Torre 3000/sala 817 CEP: 20765-000, Brazil}

\author[0000-0002-2085-9467]{M. V. Banda-Huarca}
\affiliation{Universidad Tecnológica del Perú, Arequipa, Perú}
\affiliation{Observatório Nacional/MCTI, R. General José Cristino 77, CEP 20921-400 Rio de Janeiro - RJ, Brazil}
\affiliation{Laboratório Interinstitucional de e-Astronomia - LIneA - and INCT do e-Universo. Av. Pastor Martin Luther King Jr, 126 Del Castilho, Nova América Offices, Torre 3000/sala 817 CEP: 20765-000, Brazil}

\author[0000-0001-7016-7277]{F. Marchis}
\affiliation{SETI Institute, Carl Sagan Center, 339 N. Bernardo Avenue, Suite 200, Mountain View, CA, USA}
\affiliation{Unistellar, 5 allée Marcel Leclerc, bâtiment B, Marseille, F-13008, France}

\author[0000-0002-2718-997X]{P. Tanga}
\affiliation{Université Côte d'Azur, Observatoire de la Côte d'Azur, CNRS, Laboratoire Lagrange, Bd de l'Observatoire, CS 34229, 06304, Nice Cedex 4, France}


\author{B. Blanc}
\affiliation{Unistellar, 5 allée Marcel Leclerc, bâtiment B, Marseille, F-13008, France}

\author{E. Bondoux}
\affiliation{Université Côte d'Azur, Observatoire de la Côte d'Azur, CNRS, Laboratoire Lagrange, Bd de l'Observatoire, CS 34229, 06304, Nice Cedex 4, France}

\author{A. Burnett}
\affiliation{School of Natural Resources and the Environment, University of Arizona, 1064 E Lowell St, Tucson, AZ 85721, USA}
\affiliation{Unistellar, 5 allée Marcel Leclerc, bâtiment B, Marseille, F-13008, France}

\author{S. Calavia-Belloc}
\affiliation{Sabadell Astronomical Association, Carrer Prat de la Riba, s/n, 08206, Sabadell, Catalonia, Spain}
\affiliation{Red Astronavarra Sarea de Navarra, Spain}
\affiliation{Astrosedetania de Zaragoza, Spain}

\author{O. Canales-Moreno}
\affiliation{Sabadell Astronomical Association, Carrer Prat de la Riba, s/n, 08206, Sabadell, Catalonia, Spain}
\affiliation{Red Astronavarra Sarea de Navarra, Spain}
\affiliation{Astrosedetania de Zaragoza, Spain}

\author{N. Carlson}
\affiliation{International Occultation Timing Association (IOTA), PO Box 7152, WA, 98042, USA}

\author{M. Conjat}
\affiliation{International Occultation Timing Association/European Section, Am Brombeerhag 13, 30459, Hannover, Germany}

\author{T. George}
\affiliation{International Occultation Timing Association (IOTA), PO Box 7152, WA, 98042, USA}

\author[0000-0001-6097-5297]{R. Gonçalves}
\affiliation{Polytechnic Institute of Tomar, Ci2-Smart Cities Research Center and Unidade Departamental de Matemática e Física (UDMF), 2300, Tomar, Portugal}

\author{R. Iglesias-Marzoa}
\affiliation{Centro de Estudios de Física del Cosmos de Aragón, Plaza San Juan 1, 44001, Teruel, Spain}

\author{G. Krannich}
\affiliation{International Occultation Timing Association/European Section, Am Brombeerhag 13, 30459, Hannover, Germany}

\author{D. Lafuente-Aznar}
\affiliation{Sabadell Astronomical Association, Carrer Prat de la Riba, s/n, 08206, Sabadell, Catalonia, Spain}
\affiliation{Red Astronavarra Sarea de Navarra, Spain}
\affiliation{Astrosedetania de Zaragoza, Spain}

\author{P. D. Maley}
\affiliation{International Occultation Timing Association (IOTA), PO Box 7152, WA, 98042, USA}
\affiliation{NASA Johnson Space Center Astronomical Society, Houston, TX, 77058, USA}

\author{J. Marco}
\affiliation{ASTER Agrupació Astronòmica Barcelona, Barcelona, Spain}

\author{C. Schnabel}
\affiliation{International Occultation Timing Association/European Section, Am Brombeerhag 13, 30459, Hannover, Germany}
\affiliation{Sabadell Astronomical Association, Carrer Prat de la Riba, s/n, 08206, Sabadell, Catalonia, Spain}

\author{D. Shand}

\author{K. Wagner}
\affiliation{Department of Astronomy and Steward Observatory, University of Arizona, 933 N Cherry Ave, Tucson, AZ 85719, USA}
\affiliation{Unistellar, 5 allée Marcel Leclerc, bâtiment B, Marseille, F-13008, France}
%




\begin{abstract}


Librating around the Lagrange L5, the Jupiter’s Trojan (2207) Antenor has been observed in recent years and its rotational light curve suggests it to be a very likely binary asteroid candidate. From stellar occultations, we report results from three events from Europe and North America to estimate the 2D apparent size and shape of Jupiter’s Trojan (2207) Antenor. For the best-fitted ellipse in the sky-plane, we determined that Antenor has a 2D apparent equatorial radius of $54.30 \pm 0.99$ km at the moment of the occultations, with an apparent oblateness of $0.144 \pm 0.051$. We highlight the positive detection from 2021 June 12, which shows an intriguing feature that can be interpreted as a very large topographical feature (of about 11 km) of the body or that can provide further evidence that this object is, in fact, a close or contact binary. We also determine astrometric positions, with uncertainties of a few milliarcseconds (mas) for our preferred solutions.



\end{abstract}



\section{Introduction} \label{sec:intro}

Jupiter trojans are a population of small bodies that share the giant planet's orbit around the Sun. They are grouped and librating around two equilibrium points of the circular restricted three-body problem, known as the Lagrange L4 and L5 points, in a 1:1 mean motion resonance. They are thought to be planetesimals captured from a larger population that existed in the planetary region when Jupiter was formed and/or migrated \citep[see] []{2005Natur.435..462M, 2013ApJ...768...45N, 2023SSRv..219...83B}. The first observations of Jupiter's trojans were between 1906 and 1907, with the observations of three small bodies, later named Achilles, Patroclus, and Hektor \citep{1961ASPL....8..239N}. Since then, more than $7,500$ Jupiter Trojans and more than $5,800$ candidates are currently known\footnote{Minor Planet Center Website: \url{https://www.minorplanetcenter.net/}, last checked on May/2024}. 


These small bodies can provide key details on the history and dynamic evolution of our solar system. Their composition and physical properties, such as their size and shape, can provide insights into the condition of the solar nebula in the region where they formed. Discovered in 1977, (2207) Antenor is among the largest Jupiter Trojans and is orbiting the Lagrangian point L5\footnote{Minor Planet Center - List of Jupiter Trojans: \url{https://www.minorplanetcenter.net/iau/lists/JupiterTrojans.html/}}. Its rotational light curve has been observed several times over the years, showing a rotational period near $7.964 \pm 0.001$ hours, an amplitude varying from 0.19 to 0.24
mag \citep{2011AJ....141..170M, 2017MPBu...44..252S, 2019MPBu...46..315S, 2018MPBu...45..341S} and a proper frequency of –15.69 arcsec per year \citep{2023A&A...679A..56H}. Also, \citet{2018MPBu...45..341S} found deviations from the light curve suggestive of mutual events from a binary system. Thus, more investigations about its size and shape become relevant to better constrain its physical properties.

In this context, stellar occultation is a powerful technique for studying the sizes and shapes of small bodies in our solar system \citep{1979ARA&A..17..445E}. A stellar occultation happens when one celestial body blocks the light of a background star. During this event, the shadow of the body can be observed as a sudden decrease in the measured light flux, the combination of the stellar flux and, in this case, the Trojan asteroid flux. By considering the relative velocity of the observer with respect to the body, the observation can be converted into a path on the sky, usually called \textit{chord} in the sky plane \citep{2022MNRAS.511.1167G, 2024A&ARv..32....6S}.

This method can provide sizes and shapes down to km-level accuracy, reveal atmospheres, detect features like jets, satellites, and rings of the occulting body \citep{2011Natur.478..493S, 2012Natur.491..566O, 2014Natur.508...72B, 2017AJ....154...22D, 2023Natur.614..239M}. Some previous works that have presented results for the physical properties of Jupiter Trojans from the stellar occultation technique can be found for (1437) Diomedes \citep{2000Icar..145...25S}, (617) Patroclus and its nearly equal size moon \citep{2015AJ....149..113B}, (4709) Ennomos and (911) Agamemnon \citep[][]{2014M&PS...49..103S}. For this last one, the stellar occultation technique  also revealed the presence of a satellite \citep{2013P&SS...87...78T}.

In this paper, we present three stellar occultations by (2207) Antenor. The first one was on 2021 June 12, with one positive detection  from Germany and two negatives from Spain. Then later, two multi-chord events were observed on 2021 July 10, with positive detections in France, Spain and Portugal, and the other one on 2021 August 26 from the United States and Canada. 

\section{Predictions and Observations} \label{sec:Pred}

The prediction of the stellar occultation requires the knowledge of the position of the occulted star and the ephemeris of the occulting body. For the specific events present here, predictions were made within the
framework of the European Research Council (ERC) Lucky Star project\footnote{Lucky Star Project Website: \url{https://lesia.obspm.fr/lucky-star/}}.  The campaign was managed by the Occultation Portal
website\footnote{Occultation Portal Website \url{https://occultation.tug.tubitak.gov.tr/}} described by \citet{2022MNRAS.515.1346K}. Stellar positions are from the Gaia EDR3 \citep{2022yCat.1357....0G}. Antenor's ephemeris was calculate,d and is regularly updated, using the Numerical Integration of the Motion of an Asteroid (\textsf{NIMA}) tool, which has been successfully used to determine the orbits of small bodies, described by \citet{2015A&A...584A..96D}, combined with astrometric observations carried out by various ground-based observatories \textbf{--} like Pico dos Dias Observatory (OPD\footnote{MPC code 874}) and Haute-Provence Observatory (OHP\footnote{MPC code 511}) -- and reduced with the PRAIA astrometry tool \citep{2023P&SS..23805801A}. 

Details about the observed occultations, like the date and time of the occultation and star parameters, can be found in table~\ref{tab:events}, whereas the observation circumstances for these events like information of the site coordinates, details about the telescope and camera used and whether the detection of the occultation was positive or not, are shown in table~\ref{tab:table1}. 


\begin{table*}[h!]
    \centering
    \caption{Occulted stars parameters for each observed event as obtained from Gaia DR3.}
    \resizebox{\linewidth}{!}{
    \begin{tabular}{ccccccccc}
    \hline
    \hline
    Occ. date and time & Gaia EDR3          & Right Ascension$^1$ &  Declination$^1$ & Star's & Antenor's & Event Velocity & Fresnel Scale & Star Diam. \\
    (UTC)              & source identifier  & (hh:mm:ss $\pm$ mas) & (dd:mm:ss $\pm$ mas)       &  G mag   &  V mag & (km/s) &  (km) & (km) \\ 
    \hline
    \hline
    2021-06-12 01:49:42.98 & 4100200978834179712 & 18$^h$ 45$^m$ 29$^s$.6986 ($\pm$ 0.019)  & -15$^\circ$ 38' 55''.771 ($\pm$ 0.020)  & 13.889 & 16.9 & 14.6  & 0.47 & 0.24 \\
    2021-07-10 00:56:52.52 & 4103087299953614720 & 18$^h$ 31$^m$ 04$^s$.3264 ($\pm$ 0.056)  & -15$^\circ$ 58' 45''.565 ($\pm$ 0.053)  & 12.608 & 16.9 & 16.0  & 0.47 & 2.49 \\
    2021-08-26 04:01:15.58 & 4097470749676146048 & 18$^h$ 15$^m$ 46$^s$.3889 ($\pm$ 0.016)  & -17$^\circ$ 02' 49''.535 ($\pm$ 0.016)  & 12.016 & 16.6 &  \phantom{0}4.0 & 0.49 & 0.23\\
    \hline
    \hline
   \multicolumn{9}{l}{\rule{0pt}{3.0ex} $^1$ The coordinates of the stars (RA and Dec.) along with their associated uncertainties were propagated to the occultation epoch using the formalism suggested by \cite{Butkevich_2014}}\\
    \multicolumn{9}{l}{ using the parameters (proper motion, parallax, radial velocity, etc.) from Gaia EDR3 \citep{2022yCat.1357....0G}. Their uncertainties are presented in units of milliarcseconds (mas). The}\\
    \multicolumn{9}{l}{uncertainties in RA include the factor ${\rm cos}\delta$. Antenor has a V magnitude were obtained from \citep{jplhorizons}, which can be used as a reference, but it is not}\\
    \multicolumn{9}{l}{directly comparable to G magnitude. The apparent stellar diameter is projected at the distance of the occulting object.}
    
    \label{tab:events}
    \end{tabular}
    }
\end{table*} 







\begin{table*}
	\caption{Observation Circumstances for Antenor's occultation Event}
        \resizebox{\linewidth}{!}{
	\label{tab:table1}
	\begin{tabular}{cccccc}

\hline \hline

\textbf{Site}	&	\textbf{Observer}	&	\begin{tabular}{c}
\textbf{Longitude} \\ \textbf{Latitude} \\ \textbf{Altitude (m)} \end{tabular}	&	\begin{tabular}{c} \textbf{Telescope (mm)} \\\textbf{Camera} \\ \textbf{Time Source} \end{tabular}	&	\begin{tabular}{c} \textbf{Exp. Time (s)} \\ \textbf{SNR} \end{tabular}	&	\begin{tabular}{c} \textbf{Detection} \\ 	\textbf{Sky Condition} \\ \textbf{Wind} \end{tabular}\\

\hline \hline\\
\vspace{0.05em}\\
\multicolumn{6}{c}{\textbf{2021-06-12}}\\
\hline\hline
Kaufering, Germany	&	Gregor Krannich	&	\begin{tabular}{c} $48^\circ 05' 22.7793''$ N \\ $10^\circ 50' 57.9958''$ E \\ 596 	\end{tabular} &	\begin{tabular}{c} 350 \\ QHY174M-GPS \\  Camera sync to GPS \end{tabular} & \begin{tabular}{c} 0.650 \\ 4.00 \end{tabular}	&	\begin{tabular}{c} Positive \\  Clear \\ Calm \end{tabular} \\
\hline
Barcelona, Spain	&	Jordi Marco	&	\begin{tabular}{c} $41^\circ  23' 57.997''$ N \\ $02^\circ  09' 40.0223''$ E \\ 77 	\end{tabular} &	\begin{tabular}{c}  235 \\ ZWO ASI120MM \\ Computer sync to NTP \end{tabular} & \begin{tabular}{c} 1.175 \\ na \end{tabular} &	\begin{tabular}{c} Negative \\ Clear \\ light air \end{tabular}\\
\hline
Barcelona, Spain	&	 Carles Schnabel &	\begin{tabular}{c} $41^\circ   29' 36.996''$ N \\ $01^\circ   52'  21''$ E \\ 180 	\end{tabular} &	\begin{tabular}{c} 400 \\ WATEC-910HX \\ IOTA-VTI \end{tabular} & \begin{tabular}{c} 0.160	\\  na \end{tabular} & \begin{tabular}{c}	Negative \\  Clear \\ Calm \end{tabular}\\
\hline\hline\\

\vspace{0.05em}\\
\multicolumn{6}{c}{\textbf{2021-07-10}}\\
\hline\hline
\begin{tabular}{c} Observatoire de la Côte d'Azur \\ Nice, France \end{tabular} 	&	Matthieu Conjat	&	\begin{tabular}{c} $43^\circ 43' 32.9016''$ N \\ $7^\circ 17' 59.3988''$ E \\ 300.00 	\end{tabular} &	\begin{tabular}{c} 400 \\ Asi 174mm \\ Computer sync to NTP \end{tabular} & \begin{tabular}{c} 0.200 \\ 7.14	\end{tabular} &  \begin{tabular}{c} Positive  \\ Clear \\ Calm \end{tabular}\\
\hline
Botorrita,  Spain	&	Oscar Canales-Moreno  	&	\begin{tabular}{c} $41^\circ 29' 50.55''$ N  \\ $1^\circ 1' 15.1212''$ W \\ 403.00 	\\    \end{tabular} &	\begin{tabular}{c} 500 \\ Watec 120N \\ KIWI OSD \end{tabular} &	\begin{tabular}{c} 0.160 \\ 10.20 \end{tabular}	&	\begin{tabular}{c} Positive \\ Clear \\ Calm \end{tabular}\\
\hline

Linhaceira, Portugal	&	Rui Gonçalves	&	\begin{tabular}{c} $39^\circ 31' 21.6768''$ N\\  $8^\circ 23' 1.68''$ W \\ 90.00 \end{tabular} &	\begin{tabular}{c} 254 \\ Watec 910-HX \\ GPSBOXSPRITE \end{tabular}	&	\begin{tabular}{c} 0.320 \\ 7.63 \end{tabular}	& \begin{tabular}{c}	Positive \\  Clear \\ Calm \end{tabular} \\
\hline

\begin{tabular}{c}UniversCity \\ Caussols, France \end{tabular} &	Luana Liberato &	\begin{tabular}{c} $43^\circ 45' 9.6012''$ N \\ $6^\circ 55' 18.192''$ E  \\ 1266.25 	\end{tabular} &	\begin{tabular}{c} 500 \\ QHY 600M \\ Camera sync to GPS \end{tabular}	&	\begin{tabular}{c} 0.500 \\ 14.70 \end{tabular}	&	\begin{tabular}{c} Positive \\ Clear \\ light air \end{tabular}	 \\

\hline

\begin{tabular}{c} Javalambre Astrophysical \\ Observatory \\ Teruel, Spain \end{tabular}&	Ramón  Iglesias-Marzoa &	\begin{tabular}{c} $40^\circ 2' 30.48''$ N \\ $1^\circ 0' 58.68''$ W \\ 1957.00  \end{tabular}	&	\begin{tabular}{c} Tx40 400 \\ ProLine PL4720 \\ Computer sync to NTP \end{tabular}	&	\begin{tabular}{c} 2.000 \\ 27.78 \end{tabular} &	\begin{tabular}{c}  Negative  \\ Clear \\ light breeze \end{tabular} \\

\hline

\begin{tabular}{c} Observatori Astronòmic \\ del Parc del Garraf \\ Barcelona, Spain \end{tabular} & Carles Schnabel & \begin{tabular}{c}  $41^\circ 18' 23.796''$ N \\ $1^\circ 50' 18.024''$ W \\ 297.00  \end{tabular} & \begin{tabular}{c} 280 \\ Watec - 120N \\ IOTA-VTI \end{tabular} & \begin{tabular}{c}0.640 \\ 9.00 \end{tabular} & \begin{tabular}{c}  Negative \\ Clear \\ Calm \end{tabular}\\

\hline

\hline \hline\\
\vspace{0.05em}\\
\multicolumn{6}{c}{\textbf{2021-08-26}}\\
\hline\hline

\begin{tabular}{c} Wickenburg, Arizona \\ United States \end{tabular}	&	Paul D. Maley	&	\begin{tabular}{c} 33$^\circ 58' 46.56''$ N \\ 112$^\circ 42' 42.48''$ W \\  673.00 	\end{tabular}&	\begin{tabular}{c} 200 \\ Watec 910-HX \\ Camera sync to GPS	\end{tabular}&	\begin{tabular}{c} 0.066 \\ 7.25 \end{tabular}	&	\begin{tabular}{c} Positive \\ Clear \\ Calm \end{tabular}\\
\hline
\begin{tabular}{c} Dateland, Arizona \\ United States \end{tabular}	&	Norman Carlson	&	\begin{tabular}{c} $32^\circ 45' 25.854''$ N \\  113$^\circ 38' 26.6759''$ W \\ 117.00 \end{tabular} &	\begin{tabular} {c} 235 \\ RunCam Night Eagle Astro \\ Camera sync to GPS \end{tabular}	&	\begin{tabular}{c} 0.300 \\ 6.17 \end{tabular}	&	\begin{tabular}{c} Positive \\ Clear \\ Calm \end{tabular}\\
\hline
\begin{tabular}{c} Sentinel, Arizona \\ United States
\end{tabular} &	\begin{tabular}{c} Kevin Wagner \\ Alexandra Burnett \\ Benoit Blanc
\end{tabular}	&	\begin{tabular}{c}
$32^\circ 49' 32.5''$ N \\ $113^\circ 12'  26.8''$ W	\\ 227 \end{tabular} &	\begin{tabular}{c} eVscope 1 114 \\ IMX224 \\  Sync to NTP  \end{tabular}		&	\begin{tabular}{c} 0.300 \\ 6.85 \end{tabular}	&	 \begin{tabular}{c} Positive \\ Clear \\ Calm \end{tabular} \\
\hline
\begin{tabular}{c} Kelsey, Manitoba \\ Canada \end{tabular}	&	Delwin Shand	&	\begin{tabular}{c}
$53^\circ 49' 18.4''$ N \\ $101^\circ 16' 34.7''$ W	\end{tabular}	&	\begin{tabular}{c} eVscope 1 114 \\ IMX224 \\ Sync to NTP \end{tabular}		&	\begin{tabular}{c} 0.300 \\ 4.71 \end{tabular}	&	\begin{tabular}{c} Positive \\ Clear \\ Calm \end{tabular}\\
\hline
\begin{tabular}{c} Scottsdale, Arizona \\ United States \end{tabular} &	Tony George	&	\begin{tabular}{c} $33^\circ 49' 0.0984''$ N \\ $111^\circ 52' 7.2984''$ W  \\ 843.00 \end{tabular} &	   \begin{tabular}{c}
	300 \\ Watec 910-HX	\\ Sync to GPS \end{tabular}	&	\begin{tabular}{c} 0.067 \\ na \end{tabular}	&	\begin{tabular}{c} Negative \\ Clear \\ Calm \end{tabular}\\
\hline
\multicolumn{6}{l}{\rule{0pt}{3.0ex} Data for the negative detections from June 12 and August 26 were not available (na).}
	\end{tabular}
        }
\end{table*}

\section{Data Analysis and Reduction}

In this section, we provide a detailed overview of the data analysis process. A more complete description of the procedures regarding the stellar occultations technique and methodology can be found in \cite{2024A&ARv..32....6S}. Here, we begin with the acquired Flexible Image Transport System (FITS) files and proceed to determine the light curves. All observations were tracked and all video data were corrected for timing shifts. This is done by investigating the camera features to determine, for instance, if the recorded time is that of the beginning, mid-time or end of the frame. In our calculations, we standardize all frame times as those of the mid-times. From the light curves, we fit the occultation model and identify the occultation instants, as explained in subsection~\ref{sc:photometry}. Finally, in subsection~\ref{sc:limb}, we describe how we projected the obtained times onto the tangent plane and fitted Antenor’s limb.

\subsection{Photometry and Occultation Light Curves} \label{sc:photometry}

The data sets for the events described were analyzed using the standard photometry procedures of the Platform for Reduction of Astronomical Images Automatically \citep[\textsf{PRAIA},][]{2023P&SS..23905816A}, through differential aperture photometry. The flux of the occulted star was corrected by the flux of the nearby reference stars for sky transparency variations and then normalized by the star plus occulted body flux. This step is made by applying a polynomial fit to the observed flux before and after the event. 

To derive the immersion time (disappearance) and the emersion time (reappearance) of the occulted star by the main body we used the Stellar Occultation Reduction and Analysis package \citep[\textsf{SORA}, ][]{2022MNRAS.511.1167G}, fitting for each light curve a sharp edge occultation model, which includes Fresnel diffraction, the CCD bandwidth, the stellar apparent size in kilometers at the distance of the object, and the integration time of each image~\citep[see][and references therein]{2013ApJ...773...26B, 2009Icar..199..458W}. The Fresnel scale is given by $f = \sqrt{\lambda D_o/2}$ for the diffraction observed
in the wavelength $\lambda$ from a distance to body $D_o$. For a typical
observation in the visible light $\lambda$ is $\sim 550$ nm. The values for the Fresnel scale and the star apparent diameter at the distance of the object are presented in table~\ref{tab:events}. The fitting procedure was done using a standard $\chi^2$ minimization procedure, where an adequate fitting is obtained when $\chi^2$ per degree of freedom ($\chi^2_ {pdf}$) is close to the unity. The obtained immersion and emersion times and their error bars are shown in table~\ref{tab:table3} together with the chord's size, the minimal $\chi^2_{pdf}$ obtained and light curve standard deviation (noise; $\sigma_{LC}$) in normalized flux units. Also, as a test, we reanalyzed two curves from the JUL/10 event (Linhaceira and Nice, see table~\ref{tab:table1}) without including Fresnel diffraction in the model. The differences found for the respective immersion and emersion times given by table~\ref{tab:table3} are two times smaller than their uncertainties.

\subsection{Geometric Reconstruction} \label{sc:limb}

The immersion and emersion times were projected onto the sky plane into chords, and their extremities were used to fit the shape of Antenor’s limb at the moment of the occultations. The chord’s accuracy depends primarily on the precision of the times obtained from the light curve. To date, no 3D-shape has been published for Antenor. So, as a first approximation, we assume that the occulting body limb at the moment of both occultations can be approximated by an elliptical shape and described by five parameters: the center coordinates ($f_0, g_0$) for each occultation which measures RA and DEC offsets, the apparent equatorial radius ($a'$), the apparent oblateness ($\epsilon$) and the position angle of the pole ($P$) \citep{2022MNRAS.511.1167G, 2021A&A...652A.141M}.

As will be shown in Section~\ref{sec:result}, Antenor fortuitously presented almost the same rotational phase in the observations of the July and August events, so we combined those two multi-chords occultations by fitting them with 7 varying parameters (the elliptical shape plus one center for each occultation). The event observed in June is handled separately.

The fitted parameters and their 1-$\sigma$ uncertainties are determined through minimizing the reduced $\chi_2$ via a Monte Carlo approach. For this, we used an algorithm based on the description presented by \cite{2022MNRAS.511.1167G}.

\section{Results} \label{sec:result}

In this section, we detail our results. In subsection \ref{sc:multichords} we present the determination of Antenor's 2D limb size and shape as seen in the sky-plane at the moment of two multi-chord occultations (2021 July 10 and 2021 August 26). Furthermore, in subsection~\ref{sc:singlechord}, we present the results of the June 2021 occultation. Despite being observed by only one observer, it may contain an intriguing feature suggesting that Antenor can be, actually, a close or contact binary.

\begin{table}[h!]
    \centering
	\caption{Fitted times obtained for each light curve with a positive detection.}
        	\label{tab:table3}
	\begin{tabular}{cccccc}\hline
\multicolumn{6}{c}{\textbf{2021 June 12}}\\
\hline
Site & Immersion (UT) & Emersion (UT) & Chord's size (km) &  $\chi^2_{pdf}$ & $\sigma_{LC}$ \\
\hline
Kaufering, Germany,        & 01:47:56.36 $\pm$ 0.13 & 01:48:02.37  $\pm$ 0.16 & \phantom{0}89.93  $\pm$ 3.30 & 1.25 & 0.232 \\ 
Kaufering, Germany, Seg. 1 & 01:47:56.36 $\pm$ 0.13 & 01:47:58.74  $\pm$ 0.14 & \phantom{0}35.56  $\pm$ 2.88 & 0.76 & 0.232 \\ 
Kaufering, Germany, Seg. 2 & 01:47:59.50 $\pm$ 0.14 & 01:48:02.37  $\pm$ 0.16 & \phantom{0}42.77  $\pm$ 3.21 & 0.76 & 0.232 \\ 
\hline\hline
\multicolumn{6}{c}{\textbf{2021 July 10}} \\
\hline
Nice, France         & 00:53:35.10  $\pm$ 0.03 & 00:53:40.53  $\pm$ 0.03 &  \phantom{0}88.29  $\pm$ 0.66 & 0.97 & 0.140 \\ 
Botorrita, Spain     & 00:54:07.86  $\pm$ 0.02 & 00:54:14.18  $\pm$ 0.03 & 105.90  $\pm$ 0.53            & 1.04 & 0.098 \\ 
Linhaceira, Portugal & 00:54:41.54  $\pm$ 0.04 & 00:54:48.26  $\pm$ 0.04 & 109.58  $\pm$ 1.00            & 0.96 & 0.131 \\ 
Caussols, France     & 00:53:36.93  $\pm$ 0.04 & 00:53:42.58  $\pm$ 0.04 &  \phantom{0}91.95  $\pm$ 0.92 & 0.98 & 0.080 \\ 
\hline\hline
\multicolumn{6}{c}{\textbf{2021 August 26}} \\
\hline
Wickenburg, USA   & 03:42:35.97  $\pm$ 0.01 & 03:42:59.86  $\pm$ 0.01 & 100.775  $\pm$ 0.04            & 0.83 & 0.150 \\ 
Dateland, USA     & 03:43:02.87  $\pm$ 0.02 & 03:43:26.74  $\pm$ 0.02 &  \phantom{0}85.443  $\pm$ 0.09 & 0.96 & 0.162 \\ 
Sentinel, USA     & 03:42:57.50  $\pm$ 0.03 & 03:43:21.95  $\pm$ 0.05 & 103.217  $\pm$ 0.19            & 0.90 & 0.146 \\ 
Kelsey, USA       & 03:37:40.52  $\pm$ 0.06 & 03:38:04.41  $\pm$ 0.05 &  \phantom{0}99.006  $\pm$ 0.25 & 0.90 & 0.175 \\ 
\hline
\multicolumn{6}{l}{\rule{0pt}{3.0ex} Uncertainties of the immersion and emersion times are given in seconds of UT.}
    \end{tabular}

\end{table}

\subsection{Stellar Occultation on 2021 July 10 and 2021 August 26}\label{sc:multichords}

The July 10 event was recorded by six observational stations, four of them with positive detections and two with negative ones. The observation in France had the support of the UniversCity telescope, which was developed with the goal of robotically and systematically observing stellar occultations by minor planets \citep{2020A&A...641A..81F}. The event had a shadow velocity of 15.98 km/s. The August 26 event, with the three positive detections in the United States, had a shadow velocity of 4.01 km/s. Table~\ref{tab:table1} shows details about each site, the characteristics of the cameras that recorded the event, and the exposure times. 

\begin{figure}[tb]
\begin{center}
\includegraphics[width=1\textwidth, trim={0.0cm 0.0cm 0.0cm 0.0cm}, clip]{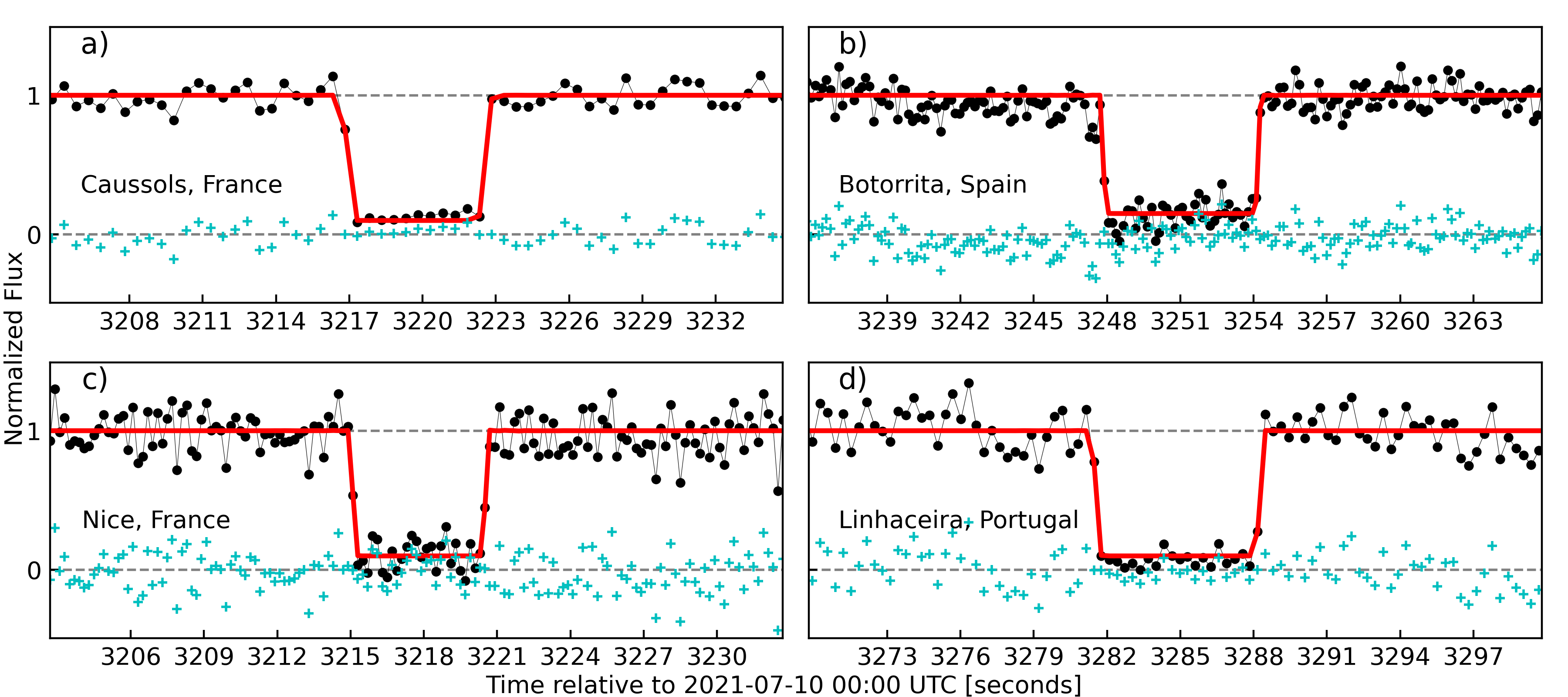}
\caption{Light curves for the occultation that occurred on 2021 July 10. The black line indicates the observed data, while the red line represents the modelled light curve. The cyan markers depict the residuals. The light curves are from (a) Caussols (France), (b) Botorrita (Spain), (c) Nice (France), and (d) Linhaceira (Portugal), with their respective observers listed in table~~\ref{tab:table1}}.
\label{fig:lc_10july}
\end{center}
\end{figure}

\begin{figure}[tb]
\begin{center}
\includegraphics[width=1\textwidth, trim={0.0cm 0.0cm 0.0cm 0.0cm}, clip]{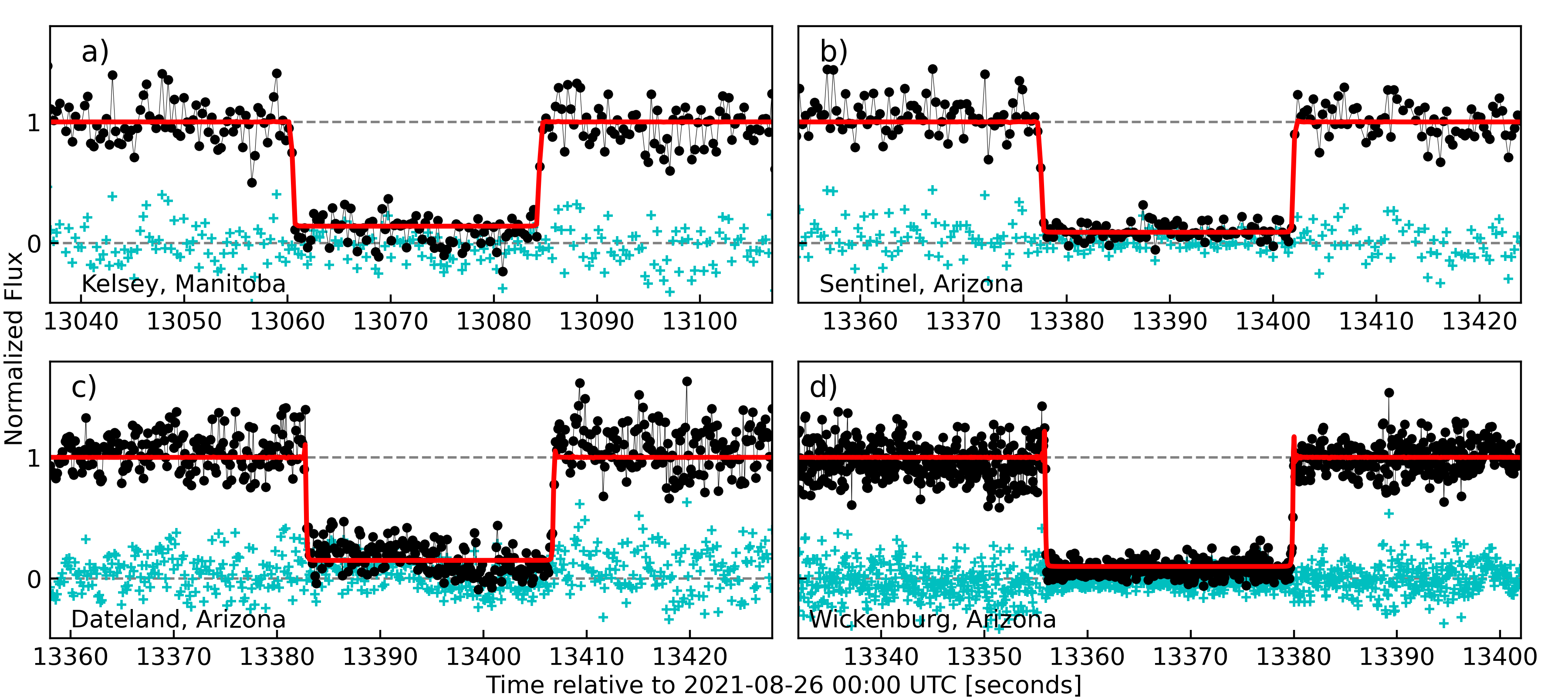}
\caption{Light curves for the  2021 August 26 occultation. The black line represents the observed data and the red line is the modeled light curve. The cyan markers are the residuals.  The light curves are from (a) Kelsey (Canada), (b) Sentinel (USA), (c) Dateland (USA), and (d)  Wickenburg (USA), whose respective observers are listed on table~\ref{tab:table1}.}
\label{fig:lc_august26}
\end{center}
\end{figure}

After normalizing the lightcurve using \textsf{PRAIA}, we obtained the immersion and emersion times  (table~\ref{tab:table3}) of the occulted stars using the procedures from the \textsf{SORA} library, as already described.  The normalized light curves for these two events are shown in figures~\ref{fig:lc_10july} and \ref{fig:lc_august26}. Also, the immersion and emersion times are used to project the chords onto the sky plane. The extremity of each chord can fit the apparent shape for its respective observation, providing the geometric reconstruction of the object, whose steps are described in subsection~\ref{sc:limb}. 

One important aspect is that considering Antenor's rotational period of $7.964 \pm 0.001$ hours \citep{ 2019MPBu...46..315S}, we compute the differential rotational phase between both multi-chords occultations (July 10 and August 26) to be around $0.016 \pm 0.018$, which means a difference of $5.8 \pm 6.5$ degrees. Considering the fact that Antenor has an orbital period of 11.76 years and, taking its position vectors on July 10 and August 26 (dates of both multi-chords stellar occultations), an angle of 3.86 (rotational phase of 0.01) degrees is found.

In addition, it should be noted that the light times on those dates differ by less than 4 minutes. Therefore, if we start counting time on the date when the object is closer, we will observe after some time a phase that is less than 4 minutes - roughly 3 degrees - later with respect to the one that would have been observed under no change in distance. The fortuitous combination of these three facts (rotational period, orbital motion, and time lag) means that we observed approximately the same projected face during both events, assuming that Antenor is not highly irregular.

\begin{figure}[tb]
\begin{center}
\includegraphics[width=1\linewidth]{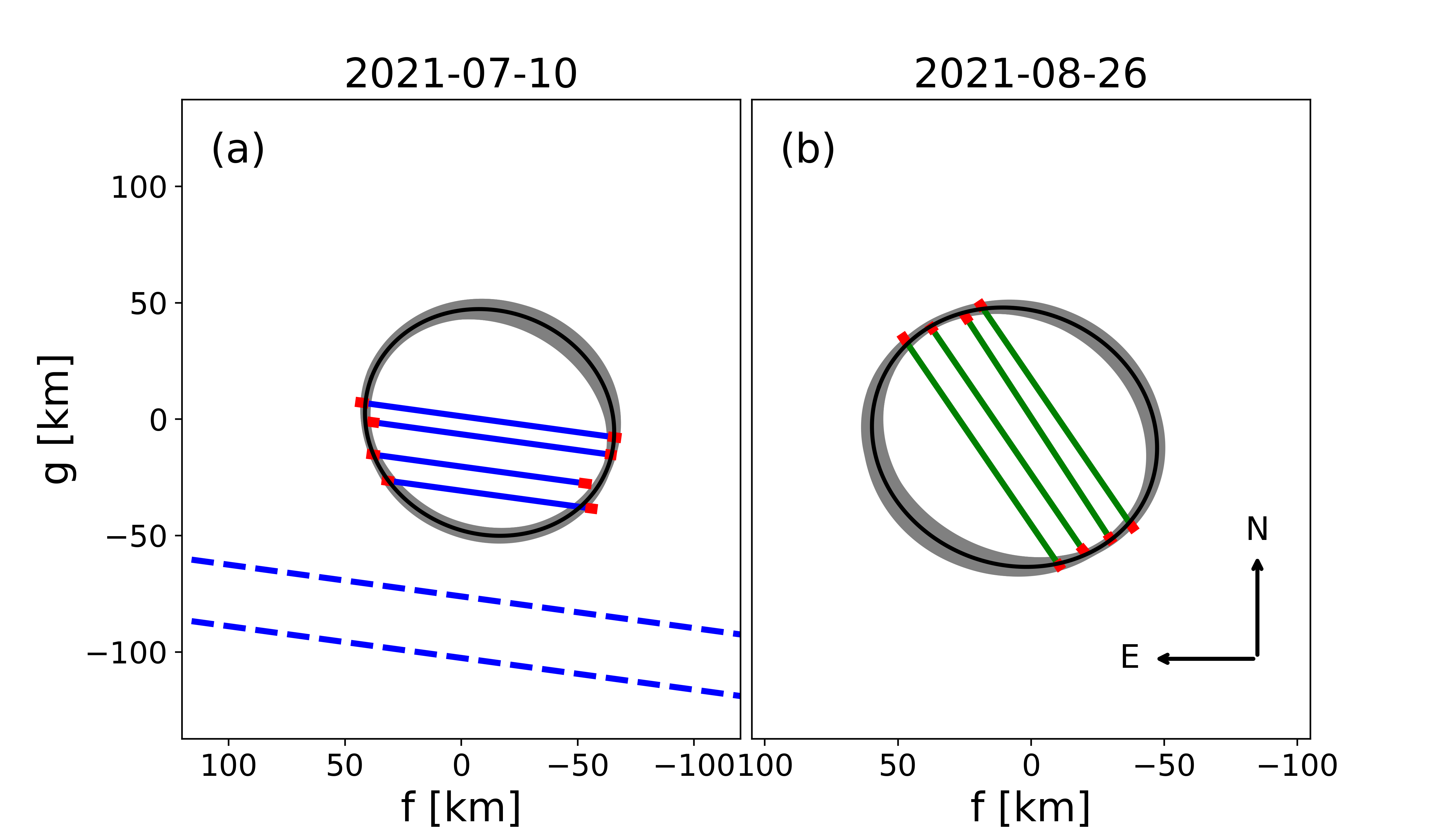}
\caption{Chords in the sky plane relative to Antenor (in blue and green), and their uncertainties (in red). The black line is the best-fitted ellipse, and all the ellipses in the 1$\sigma$ region are in gray. The left panel (a) is the event on 2021 July 10, and the right panel (b) is the event on 2021 August 26. The dashed lines stand for the negative chords of the July occultation. The data for the negative observation of August 26 was not provided by the observer.}
\label{fig:july_august}
\end{center}
\end{figure}

Figure~\ref{fig:july_august} displays the best-fitted ellipses along with their respective 1-$\sigma$ regions, representing Antenor’s limb for observations on July 10 and August 26. By analyzing both multi-chord events discussed here, we derived a global 2D apparent elliptical shape with a semi-major axis of $54.30 \pm 0.99$ km and an apparent oblateness of $0.11 \pm 0.05$. This indicates an area equivalent radius\footnote{Area equivalent radius ($R_{eq}$) is the radius of the circle whose area is equivalent to the fitted ellipse ($R_{eq} = \sqrt{a \times b}$, where $a$ is the apparent semi-major axis and $b$ is the apparent semi-minor axis), measured in kilometers \citep[see][]{2022MNRAS.511.1167G}.} of $50.86 \pm 1.13$ km. Furthermore, a radial dispersion of $2.20$ km was obtained. This dispersion may be indicative of topographical features up to this level in the radial direction apart from the obtained elliptical global shape for this body, and assuming topographic features up to this value, we obtained a minimal chi-square per degree of freedom ($\chi^{2}_{pdf}$) of 1.35 \citep[see discussion in ][]{2022MNRAS.511.1167G}. 

The physical parameters of the fitted ellipse are presented in table~\ref{tab:table4}. It is important to highlight that the fitted ellipse is used to mainly assess dimensions of the body and not its 3D shape. Results from these two occultations by Antenor are also presented by \citet{2023A&A...679A..56H}, whose determined an area equivalent of 101 $\pm$ 3 km, which is in good agreement with the 101.7 $\pm$ 2.3 km we found in the present manuscript. 

The 2021-08-26 data in \citet{2023A&A...679A..56H} show two chords (this paper shows four) and they considered the occultations individually, not combined as we did. We presented, however, this result in a separated image procedure since they show more trustworthy uncertainties based on our methodology. The combined solution 
gives a more accurate equivalent diameter as compared to that obtained from analyzing the July and August observations separately. In addition, we also presented an astrometric results using NIMA solution (see section~\ref{sc:astrometry}). It is important to highlight that the obtained area equivalent radius of $50.86 \pm 1.13$ km is also close to the value $48.8 \pm 0.2$ km published by \cite{2012ApJ...759...49G}, which used thermal data from NEOWISE.

\begin{table}[h!]
    \centering
	\caption{Physical Parameters of the fitted ellipse}
        	\label{tab:table4}
	\begin{tabular}{lcc}
\hline
\hline
&  \textbf{2021 July 10} & \textbf{2021 August 26} \\
\hline
\hline
$f_0$ (km)             &  $-12.14 \pm 1.21$            & $+6.23 \pm 2.73$ \\
$g_0$ (km)             &  $-\phantom{1}1.41 \pm 2.65$  & $-6.78 \pm 1.71$ \\
Apparent semi-major axis (km)   &  \multicolumn{2}{c}{$54.30 \pm 0.99$} \\
Apparent semi-minor axis (km)   &  \multicolumn{2}{c}{$47.91 \pm 2.15$} \\ 
Apparent equivalent radius (km) &  \multicolumn{2}{c}{$50.86 \pm 1.13$} \\ 
Apparent oblateness             &  \multicolumn{2}{c}{$\phantom{0}0.114 \pm 0.051$} \\
Position angle of the minor axis (deg)   &  \multicolumn{2}{c}{$159.3 \pm 10.3$} \\
\hline
\hline
	\end{tabular}
\end{table}

\subsection{Stellar Occultation on 2021 June 12} \label{sc:singlechord}

\begin{figure}[h!]
\begin{center}
\includegraphics[width=1.0\textwidth, trim={2.0cm 0.5cm 2.0cm 2.0cm}, clip]{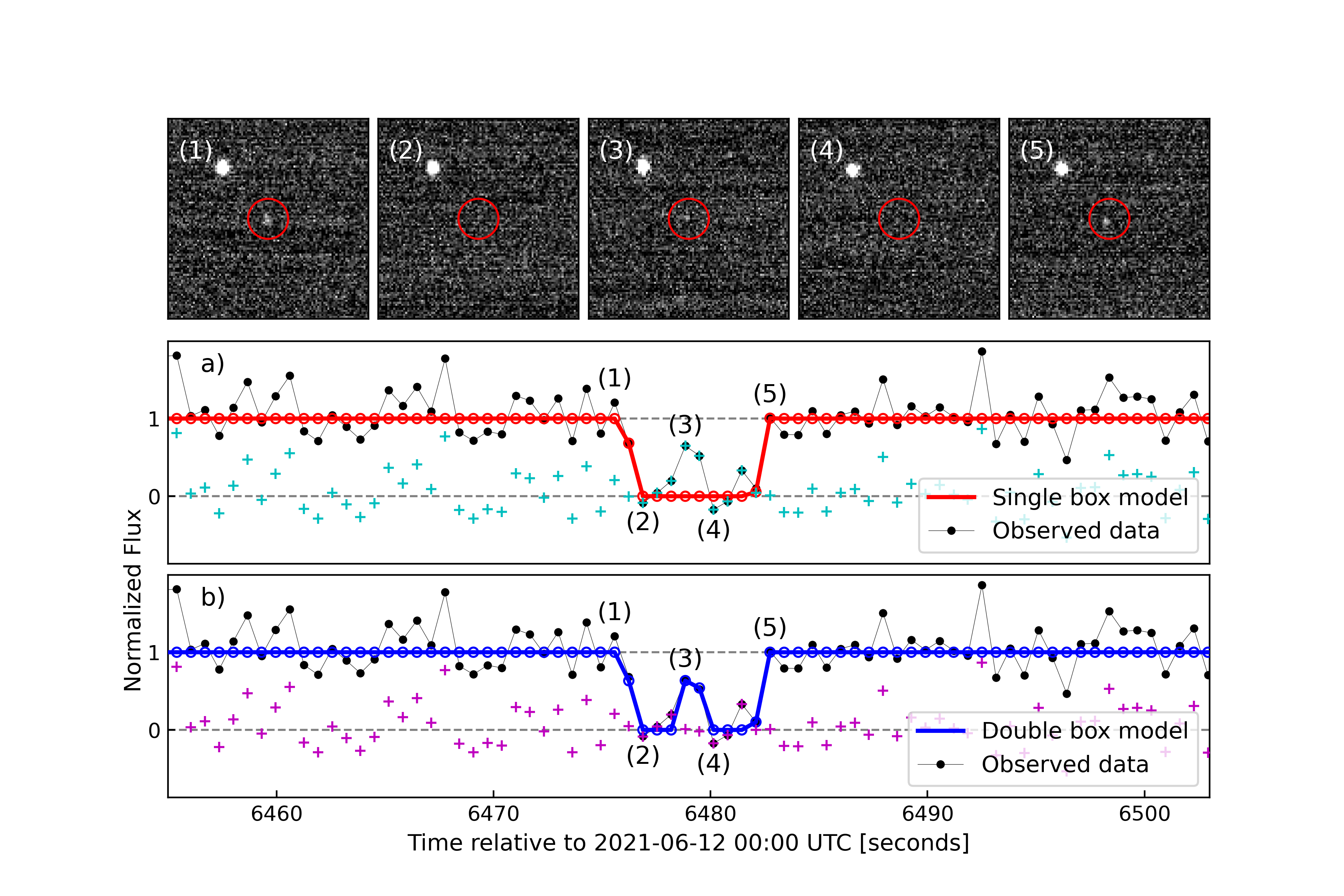}
\caption{Double-chord observed for 2021 June 12 occultation. The figure presents a sequence of images showing the moment of the occultation by Antenor in 2021 June 12. The points (1), (2), (3), (4) and (5) represents five moments of the occultation event.  The moment when the star flux increases during the occultation is represented by the point (3).The figure also shows the double drop of the observed data (black line), the single box model (red line, panel a), and the double box model (blue line, panel b). After applying a F-test statistic, it was verified that the double box model best fits the data set considering a probability of 80\%.}
\label{fig:lc_june12}
\end{center}
\end{figure}

The occultation observed on June 12 had only one positive detection with a shadow velocity of 14.63 km/s. One interesting fact about this event is a flux increase in the middle of the occultation, which can be attributed to the reappearance of the occulted star during the occultation. Figure \ref{fig:lc_june12} shows a sequence of images and points of the event, where we can see the increase of the flux of the star during the occultation (see point (3)). This kind of feature may indicate a large topographical feature of the body or binarity. Considering this peculiarity, we fitted two models: (i) a single occultation model as described in \ref{sc:photometry}; and (ii) a model containing two occultation models, both of them affected by the same effects (diffraction, stellar diameter, and instrumental response). In the first model, two parameters were fitted: the immersion and emersion times. On the other hand, the second model considered four parameters (e.g., the immersion and emersion times for each occultation).

To check whether the best model that fits our data set is the single box or double box, we applied a Fisher-Snedecor F-test to the residual of both models, which compares the variances of two samples under the null hypothesis that they result from the same distribution. Suppose the $f$ value is smaller than a critical value (considering a typical probability). In that case, we fail to disprove the null hypothesis, meaning there are no significant variations between both distributions. We only considered the points close to the occultation and, as a result, we obtained a $f$ value of 1.5288 with a critical value of 1.4451 for a probability of $80\%$. Thus, considering an 80\% probability, the second model fits the data better. Even though this is not a strong evidence (larger than 99\%, for instance), it is indicative that it is not a simple noise contamination and can be interpreted as Antenor's having a more complex shape than a simple ellipsoid can explain. A few examples of possible explanations would be: Antenor's has a large topographical feature or this object is, in fact, a close binary. However, without more and better data, it is impossible to pinpoint exactly which scenario is more likely.  Figure~\ref{fig:lc_june12} shows the light curve for this observation and also presents the fitting for the single box model (in red) and for the double box model (in blue).

The occultation instants using both models can be found in table~\ref{tab:table3}. We then projected these instants on the sky plane as chords as described in subsection~\ref{sc:limb}. Moreover, again assuming the Antenor's rotational period, we computed the differential rotational phase between this occultation and the others, and obtained a value of $0.286 \pm 0.028$ or $102.92 \pm 10.23$ degrees, which means that now we are observing a completely different face of Antenor. We estimated the length of the space between the two chords' segments and obtained the size of $11.38 \pm 2.98$ km. 

\begin{figure}[h!]
\begin{center}
\includegraphics[width=0.5\linewidth]{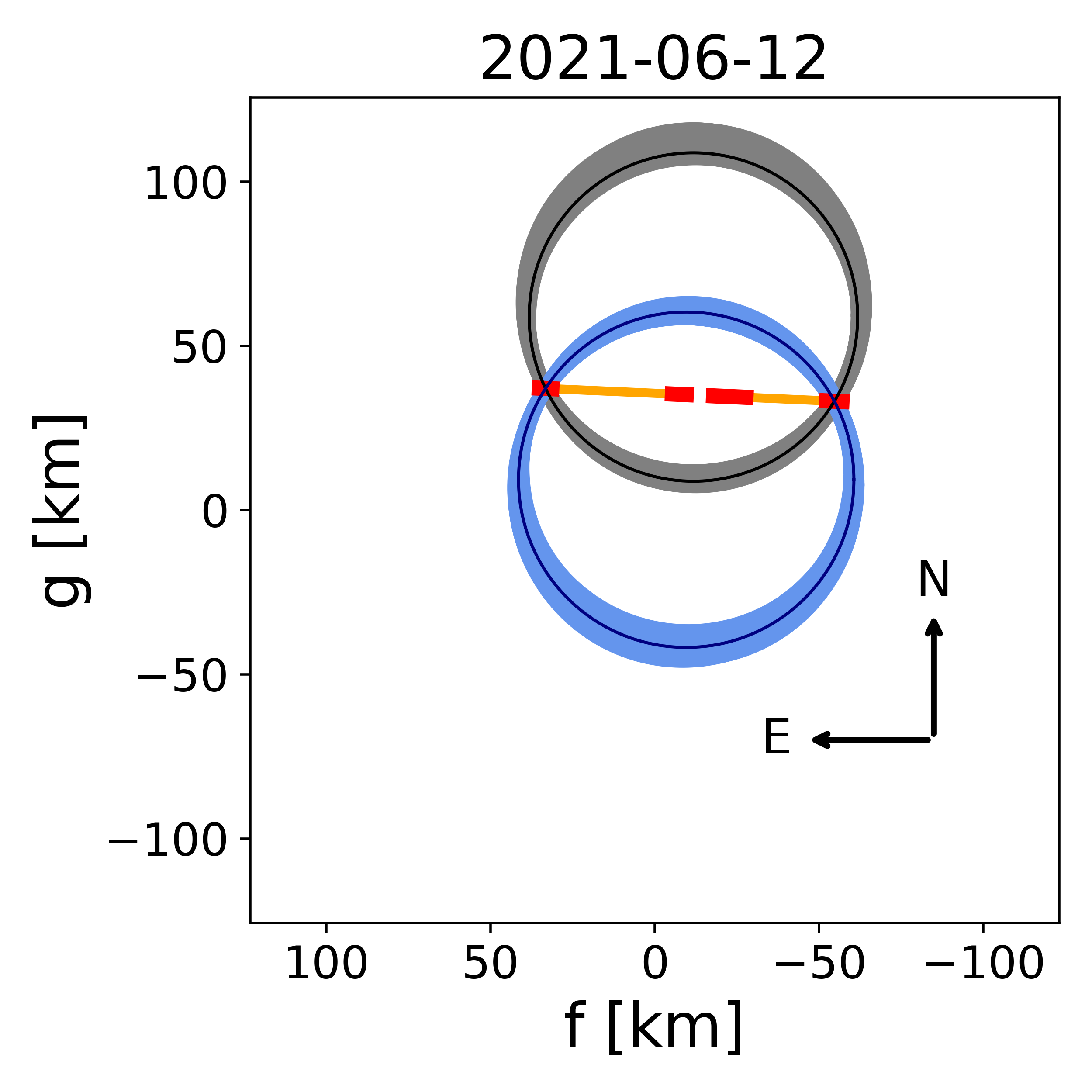}
\caption{Best fitted limb to the double-chord and ellipses within their respective $1\sigma$ region. Two solutions can be obtained, one with the centre north of the chord (in black), and another to the south (in blue). The solution in blue is the preferred solution.}
\label{fig:two_solution}
\end{center}
\end{figure}

Without further information on the object's shape, we can not pinpoint exactly where Antenor's center is, which means we can not say with certainty if this chord is near the edge or the center line. However, as a first-order estimation, we considered a circular solution of radius $50.86 \pm 1.13$ km as the value obtained in section \ref{sc:multichords}, which resulted in two possible solutions as illustrated in figure~\ref{fig:two_solution}. One solution is with the center to the North of the chord (in black), and the second with the center to the South (in blue). A more conservative approach is to combine both positions, having the mean value as the best-fitted value, with a larger uncertainty \cite[see discussion in][]{2020A&A...644A..40R}. Moreover, it is unlikely that the north solution is correct, as it presents an offset larger than the expected value of the uncertainty of Antenor's ephemeris, see details in subsection \ref{sc:astrometry}. 


\subsection{Astrometrical measurements and orbital improvement} \label{sc:astrometry}

\begin{figure}[h]
\begin{center}
\includegraphics[width=1\linewidth]{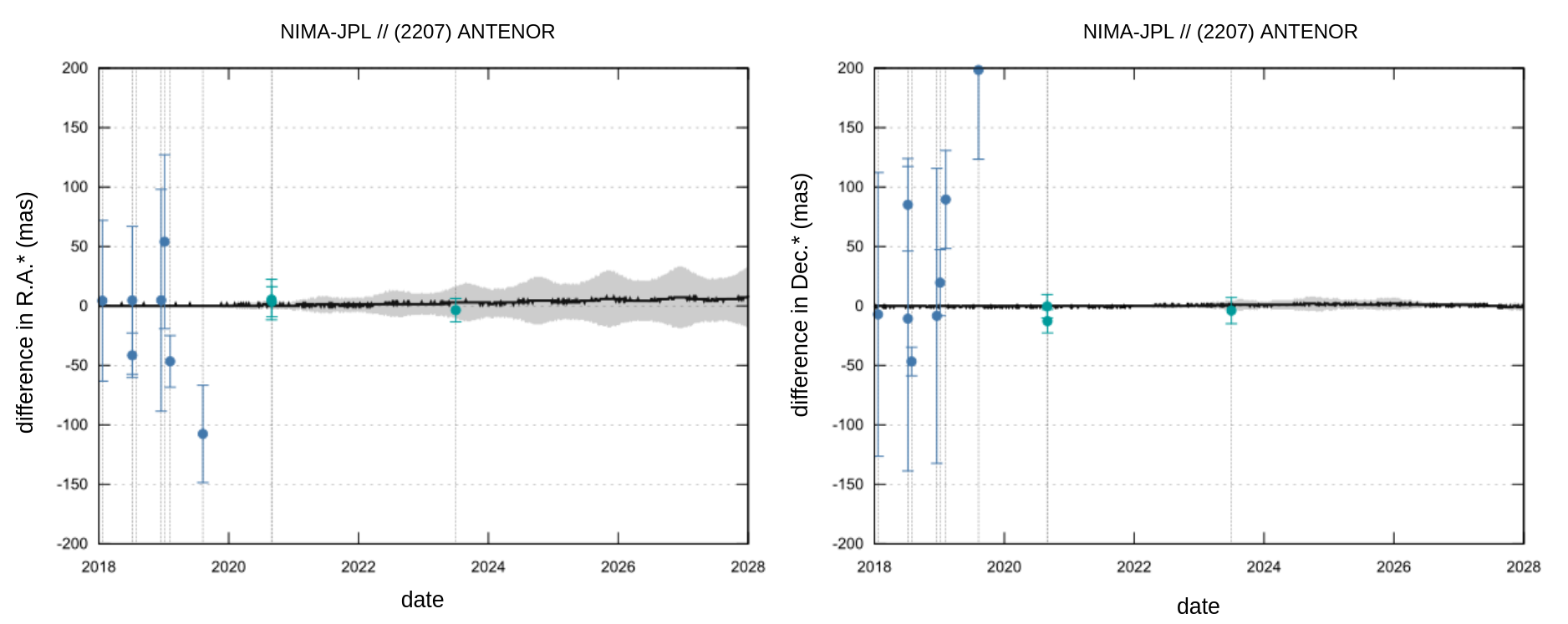}
\caption{Difference between JPL ephemeris and NIMA solution in RA (left panel) and DEC (right panel) considering all data available on MPC and proprietary astrometrical positions obtained at OPD and OHP (cyan dots), plus data from GaiaFPR (blue dots). The gray region delimits the 1-$\sigma$ uncertainty of the NIMA ephemeris.}
\label{fig:mpc_gaia}
\end{center}
\end{figure}

\begin{figure}[h!]
\begin{center}
\includegraphics[width=1\linewidth]{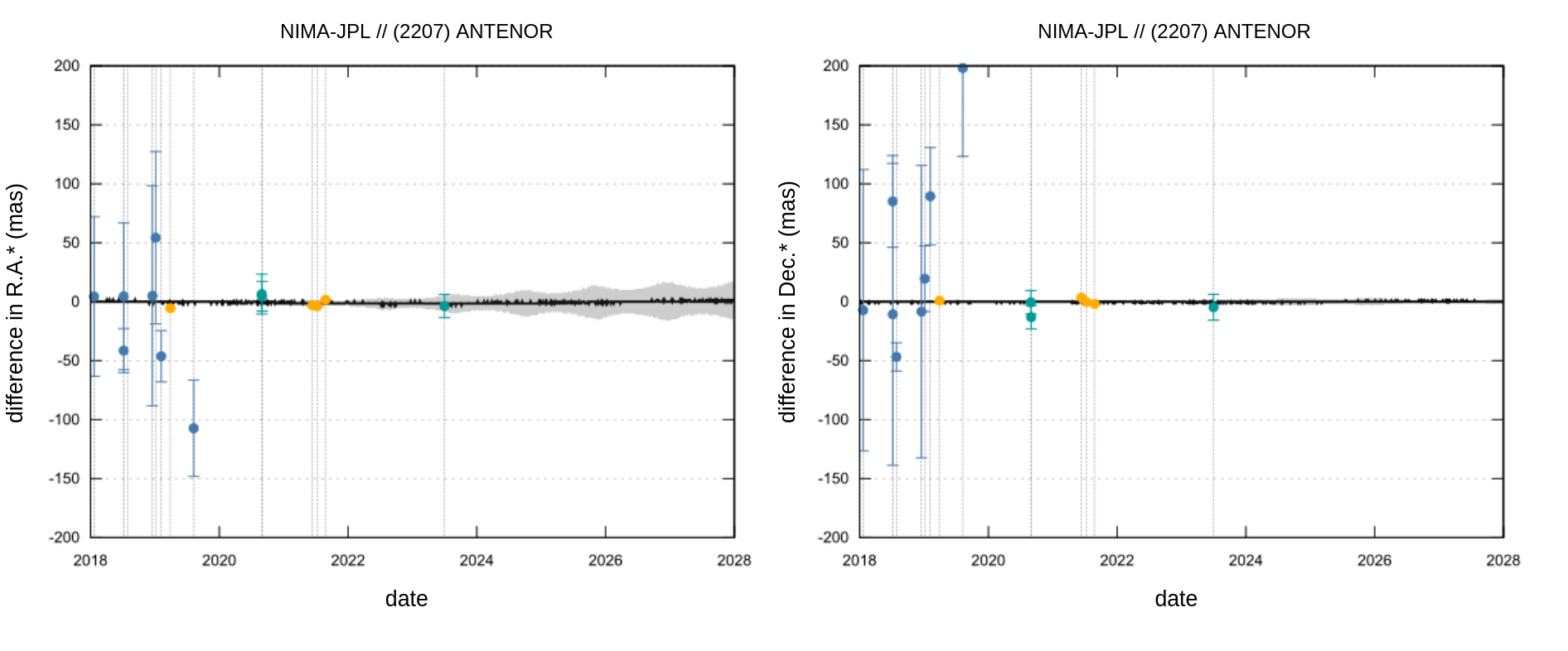}
\caption{Difference between JPL ephemeris and NIMA solution with the data presented in figure \ref{fig:mpc_gaia} plus the preferred astrometrical positions obtained in the project and presented in table \ref{tab:table5} (yellow dots). Note that the gray region that delimits the 1-$\sigma$ uncertainty of the NIMA ephemeris is smaller than the one presented in figure \ref{fig:mpc_gaia} by a factor of $\sim$2.}
\label{fig:allset1}
\end{center}
\end{figure}

A result of this technique is the geocentric astrometric position of Antenor at the occultation instants. These results are shown in table ~\ref{tab:table5}. The uncertainties are presented in milliarcseconds (mas) and those in right ascension include the multiplication by $\rm{cos}\delta$. To better understand how our data improved Antenor ephemeris, we used NIMA \citep{2015A&A...584A..96D} to fit two sets of data: the first, with all available data on MPC, and data from GaiaFPR \citep{2023LPICo2851.2209T} and proprietary astrometrical positions obtained at OPD and OHP; the second with the first set plus the preferred occultation positions presented here. Figures~\ref{fig:mpc_gaia} and \ref{fig:allset1} show these orbital fittings and the difference between the new NIMA v5 ephemeris and JPL62 solution. As mentioned earlier in the text, the astrometry for the June event is model dependent and a better constraint on Antenor's shape is expected to highly improve this result.

\begin{table}[h!]
    \centering
	\caption{Astrometric object position}
        	\label{tab:table5}
	\begin{tabular}{cccl}\hline
Date and Time (UTC) & Right Ascension (RA) & Declination (DEC) & Comment \\
\hline
 2021-06-12 01:49:42.980 & $\phantom{-}18^h 45^m 29^s.6915281 \pm 0.567$ mas & $-15^o 38' 53''.980590 \pm 1.617$ mas & Center to the North \\
 2021-06-12 01:49:42.980 & $\phantom{-}18^h 45^m 29^s.6915814 \pm 0.598$ mas & $-15^o 38' 53''.996995 \pm 1.641$ mas & Center to the South (*) \\
 2021-06-12 01:49:42.980 & $\phantom{-}18^h 45^m 29^s.6915555 \pm 0.968$ mas & $-15^o 38' 53''.988804 \pm 9.827$ mas & Combined solution \\

 2021-07-10 00:56:52.520 & $\phantom{-}18^h 31^m 04^s.3100087 \pm 0.507$ mas & $-15^o 58' 43''.998811 \pm 0.932$ mas & Same shape as 2021-08-26 \\

 2021-08-26 04:01:15.580 & $\phantom{-}18^h 15^m 46^s.3494335 \pm 0.815$ mas & $-17^o 02' 49''.230749 \pm 0.519$ mas & Same shape as 2021-07-10 \\
\hline
\multicolumn{4}{l}{\rule{0pt}{3.0ex} $(*)$ Preferred solution due to smaller offset relative to Antenor's ephemeris. The uncertainties are presented in units of milliarcseconds}\\
\multicolumn{4}{l}{\rule{0pt}{3.0ex} (mas) and those in right ascension are multiplied by $\rm{cos}\delta$.}
	\end{tabular}
\end{table}

Our orbital fitting using NIMA provided ephemeris with uncertainties below 15 mas up to 2026, which translates to 46 km projected at Antenor's distance. Without using these occultations data Antenor ephemeris would have uncertainties of 30 mas up to 2026 (92 km at Antenor's distance), which is larger than Antenor's radius of 50.83 km. This new ephemeris allows the prediction of future accurate stellar occultations with an expected high success rate. This new ephemeris is publicly available at the Lucky Star webpage\footnote{Available at \url{https://lesia.obspm.fr/lucky-star/obj.php?p=1045}}.

\section{Albedo}

The accurate diameter of Antenor, projected onto the plane of sky,
encourages the determination of its geometric albedo. However, due to its relatively short orbital period around the Sun (as compared to trans-neptunian objects, for instance), attention must be made on the dependency between the absolute magnitude and the aspect angle (angle between the asteroid's spin axis and the viewing direction) \citep[see][for more details]{2024A&A...687A..38C}.

In this context, we took\footnote{https://fink-broker.org/} a set of 14 apparent magnitudes in the filter \textit{g} and 12 apparent magnitudes in the filter \textit{r} of Antenor, obtained by the ZTF \citep[Zwicky Transient Facility][]{2019PASP..131a8003M} in June, July and August 2021. Therefore, close to the dates of the events considered here.

\begin{figure}[h]
\begin{center}
\includegraphics[width=0.5\linewidth]{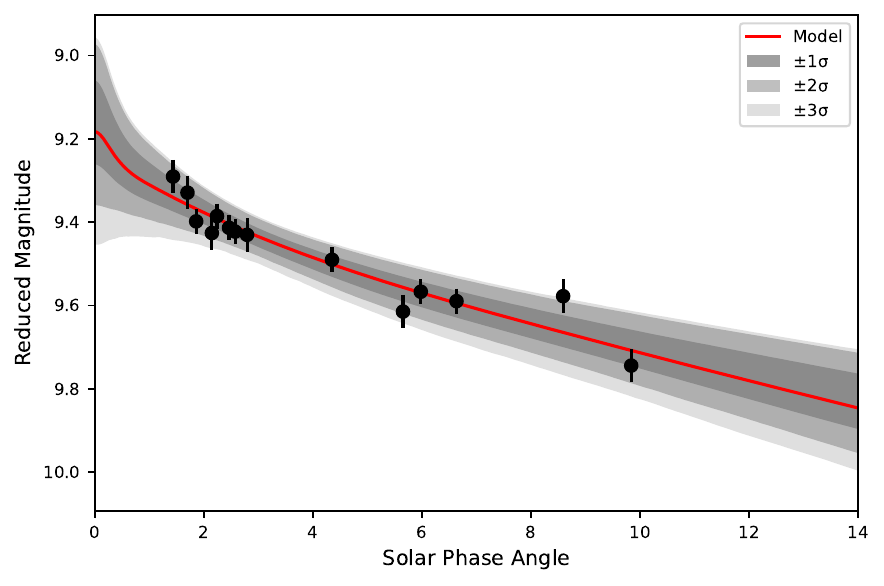}\includegraphics[width=0.5\linewidth]{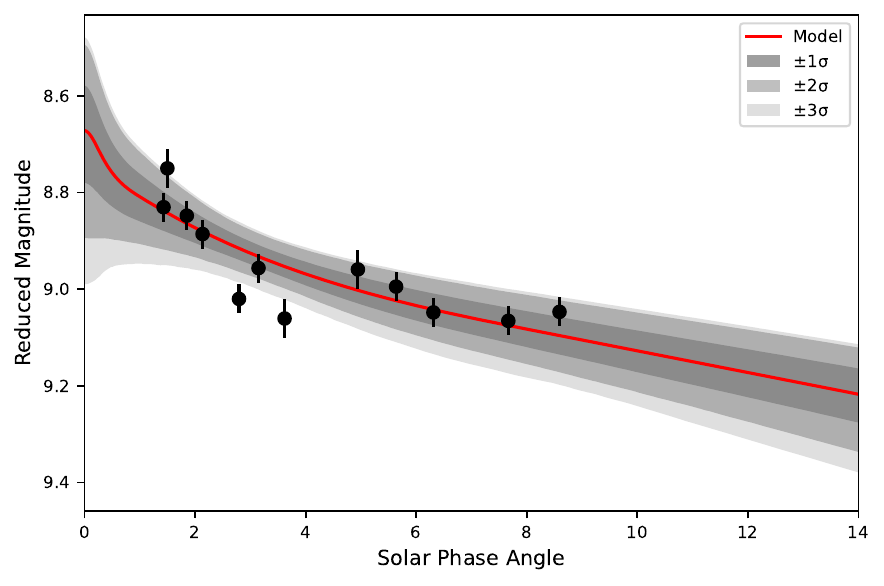}
\caption{Apparent magnitude reduced to unit distance as a function of the Solar phase angle. Left plot: \textit{g} filter. Right plot: \textit{r} filter. Photometric measurements are from the Zwicky Transient Facility. The error bars have a total length of $2\sigma$ and take into consideration the observational uncertainty in the magnitude only.}
\label{fig:abs_mags}
\end{center}
\end{figure}

Absolute magnitudes in both filters were then calculated using the \textit{HG1G2} model \citep{2010Icar..209..542M}. These values are $H_g=9.159^{+0.101}_{-0.099}$, $G1_g=0.444^{+0.118}_{-0.349}$, $G2_g=0.354^{+0.198}_{-0.128}$ and $H_r=8.674^{+0.106}_{-0.095}$, $G1_r=0.094^{+0.298}_{-0.086}$, $G2_r=0.615^{+0.112}_{-0.163}$ ($1\sigma$ confidence intervals). The model fit along with different confidence intervals are shown in Fig.~\ref{fig:abs_mags}

A rotational light curve model could not be determined from the set of ZTF measurements used here. However, we can include the effect of Antenor's rotation in the uncertainty of the absolute magnitudes in both filters by following, for instance, the procedure detailed below used by \citet{2025MNRAS.540..460F}:
\begin{itemize}
    \item For the magnitude measurements in each band (\textit{g}, \textit{r}), we determine the respective reduced magnitudes (\textit{y} axis values in both panels of Fig.~\ref{fig:abs_mags}). These are the observed apparent magnitudes normalized by Antenor's distance from the Sun and from the observer or, more specificcally, $m-5\log(R\Delta)$, where \textit{m} is the observed apparent magnitude, \textit{R} is the distance to the observer and $\Delta$ is the heliocentric distance.
    \item To each reduced magnitude, the value of the function $f(\phi)=A\times sin(\phi)$ is determined for $\phi$ randomly chosen in the interval [0:2$\pi$[, with $A=0.19$. It is clear that, here, we adopt a light curve with an amplitude of 0.19 magnitudes \citep{2011AJ....141..170M} and a sinusoidal shape. Note that a different model would only influence the uncertainties of the absolute magnitudes in each filter only and not their final values under the reasonable hypothesis that rotational phase and the ZTF observation dates are not correlated. Given the scarcity of data, the choice for a simple shape model (sine wave in our case) is a sensible choice.
    \item To fully account for the variance in the ZTF data, magnitudes used in the simulation are random values sampled from a normal distribution centered on the reduced magnitude value, with a variance corresponding to the uncertainty of the apparent magnitude ($\pm 1\sigma$).
    \item Values of $H_{g,r}$, $G1_{g,r}$ and $G2_{g,r}$ were determined using least squares minimization applied to \textit{HG1G2} phase curve models, through a constrained optimization linear approximation algorithm \citep{powell1994cobyla}.
    \item Steps 2, 3 and 4 are repeated 10\,000 times.
    \item The final values and uncertainties of $H_{g,r}$, $G1$ and $G2$ are derived from the distribution of parameters obtained from all simulations, with uncertainties determined by the $16^{\rm th}$ and $84^{\rm th}$ percentiles, representing the interval corresponding to $\pm 1\sigma$.
\end{itemize}

The determination of the geometric albedo for Antenor, in the ZTF \textit{g} and \textit{r} bands, can be readily determined from the relation \citep[e.g.][]{1916ApJ....43..173R,1997Icar..126..450H}

\begin{equation}
    \rho_{g,r}=\left(\frac{2\rm{AU}\times10^{0.2(m_\odot-H_{g,r})}}{D}\right)^2
\end{equation}

\noindent where $\rho_{g,r}$ is the geometric albedo in the \textit{g} or \textit{r} bands, $m_\odot$ is the apparent magnitude of the Sun in the respective band, $H_{g,r}$ is the absolute magnitude of Antenor in the \textit{g} or \textit{r} bands and \textit{D} is the equivalent diameter from the occultation. AU is the Astronomical Unit in the same unit of the diameter.

The ZTF image processing pipeline calibrates its photometry \citep[][]{2019PASP..131a8002B} to PanSTARRS1 \citep[Panoramic Survey Telescope and Rapid Response System,][]{2016arXiv161205560C}. The PanSTARRS1 implements the AB magnitude system with an accuracy of $\sim 0.02$ mag (90\% confidence) \citep{2012ApJ...750...99T} so we adopt from \citet{2018ApJS..236...47W} the AB values of the apparent magnitude of the Sun for PanSTARRS1 in the \textit{g} band (PS1\_g) $m_\odot = -26.54$ and in the \textit{r} band (PS1\_r) $m_\odot = -26.93$, both values with an accuracy of 2\%. The geometric albedos of Antenor are, as a consequence, $\rho_g = 0.046^{+0.008}_{-0.007}$ and  $\rho_r = 0.050^{+0.008}_{-0.007}$. Differences between these albedos and those given by other works \citep[e.g.][]{2003AJ....126.1563F} are due to different values for the diameter of Antenor mainly.

\section{Conclusions and Discussion}

Stellar occultation by Solar System Objects is an effective technique to obtain the physical characteristics of these bodies. Networks of ground-based telescopes have been used to observe such events and obtain as many chords as possible that can be combined to reveal the size and shape of small bodies. In this context, we used three observations of a stellar occultation by the Trojan (2207) Antenor and determined its physical characteristics. These results contribute to a better understanding of this object's physical properties and can provide insights into the Jupiter Trojans population.

Two multi-chords occultations in 2021 July 10 and 2021 August 26, allowed the characterization of Antenor's limb at the moment of the occultations. During those events, Antenor can be considered at almost the same rotational phase. Observations were made by professional and citizen astronomers in Europe and the United States. From these events, we estimated that Antenor's shape projected onto the plane of the sky can be fitted by an ellipse with a semi-major axis of $54.30 \pm 0.99$ km and semi-minor axis of $47.91 \pm 2.15$ km, which means an oblateness of $0.114 \pm 0.051$. From these values, we computed the area equivalent radius of $50.86 \pm 1.13$ km. Our result can be compared with $48.2 \pm 0.2$ km which is the expected value obtained by thermal measurements using the NEOWISE \citep{2012ApJ...759...49G}. Results from these two events are also presented by \citet{2023A&A...679A..56H}. Our analysis, however, follow a different methodology and combine chords from different events, taking into consideration that Antenor profited from almost the same rotational phase. In addition, we provided useful and accurate astrometric results using the NIMA solution.

Considering the Antenor's absolute magnitudes in the \textit{g} and \textit{r} bands for observations close to the dates of the events analyzed in this work, we computed the geometric albedo in both filters  using the area equivalent radius obtained in the paper. The obtained values are $\rho_g = 0.046^{+0.008}_{-0.007}$ and  $\rho_r = 0.050^{+0.008}_{-0.007}$. The results found here for Antenor’s albedo are in good agreement with the values of $\rho_V = 0.051 \pm 0.003$ and $\rho_V = 0.059 \pm 0.003$ presented by NEOWISE \citep{2012ApJ...759...49G} and Akari \citep{2011PASJ...63.1117U} respectively. With a low-albedo, Antenor is classified as a D-type asteroid \citep{2011AJ....141..170M}, which is the taxonomy that shows predominance for L5 trojans \citep{2008A&A...483..911R}. 


The 2021 June 12 occultation presented only one positive chord detected by Gregor Krannich at Kaufering, Germany. This light curve had a two-point increase in the light flux in the middle of the occultation which is more significant than the 1-$\sigma$ noise of the light curve. This can be interpreted as a reappearance of the occulted star in the middle of the event. Moreover, this indicates a large topographic feature or that Antenor is a contact or close binary. The unocculted region in the middle of the occultation has a size of $11.38 \pm 2.98 $ km, a very large ($22$\%) value considering the $50.86$ km area equivalent radius determined. Moreover, our first-order estimation puts this chord at a distance of $25.43$ km from Antenor's center, which agrees with a binary object as suggested by \citet{2018MPBu...45..341S}. More observational data is needed to constrain this scenario better and pinpoint the individual sizes of the components.    

From these occultations, we determined the relative position between Antenor and the occulted star. Based on precise knowledge of the star position due to the Gaia catalog, we determined Antenor's positions in the ICRS with uncertainties between 0.5 and 1.6 mas. These new positions were used to improve Antenor's orbital parameters, allowing the prediction of the path of future stellar occultations with a high accuracy (uncertainties smaller than 15 mas), which is smaller than Antenor's radius. 

Out of approximately $\sim7,500$ known Trojan objects\footnote{Source: IAU Minor Planet Center, last checked on May/2024}, only $\sim$1,800 have their sizes estimated from thermal measurements \citep{2012ApJ...759...49G}. A small number of objects have been measured by occultations\footnote{Source: \url{https://occultations.ct.utfpr.edu.br/results/}} \citep{2019JPhCS1365a2024B}, and even fewer have been detected in multi-chord occultations. Accurate sizes and shapes of these objects contributes to advance our knowledge of the formation and dynamical evolution of the Solar System.


\section*{}
We acknowledge a CAPES support. This study was financed in part by Coordenação de Aperfeiçoamento de Pessoal de Nível Superior - Brazil (CAPES) - Finance Code 001, also by CAPES-PRINT Process 88887.570251/2020-00, by the French Programme National de Planetologie, and by the BQR program of Observatoire de la Côte d'Azur. JIBC acknowledges grants 305917/2019-6, 306691/2022-1(CNPq) and 201.681/2019 (Rio de Janeiro State Research Support Foundation, FAPERJ). FBR acknowledges CNPq grant 316604/2023-2. Partially based on observations made with the Tx40 telescope at the Observatorio Astrofísico de Javalambre in Teruel, a Spanish Infraestructura Cientifico-Técnica Singular (ICTS) owned, managed, and operated by the Centro de Estudios de Física del Cosmos de Aragón (CEFCA). Tx40 is funded with the Fondos de Inversiones de Teruel (FITE).

Based on observations obtained with the Samuel Oschin Telescope 48-inch and the 60-inch Telescope at the Palomar Observatory as part of the Zwicky Transient Facility project. ZTF is supported by the National Science Foundation under GrantNo. AST-2034437 and a collaboration including Caltech, IPAC, the Weizmann Institute for Science, the Oskar Klein Center atStockholm University, the University of Maryland, Deutsches Elektronen-Synchrotron and Humboldt University, the TANGOConsortium of Taiwan, the University of Wisconsin at Milwaukee, Trinity College Dublin, Lawrence Livermore NationalLaboratories, and IN2P3, France. Operations are conducted by COO, IPAC, and UW.

%





\bibliography{sample631}{}
\bibliographystyle{aasjournal}



\end{document}